\documentclass[journal, compsoc]{IEEEtran}

\usepackage[bigdelims,vvarbb,varg]{newpxmath}

\IEEEoverridecommandlockouts 

\usepackage[style=ieee]{biblatex}
\DefineBibliographyExtras{english}{%
    \DeclareBibstringSet{latin}{andothers,ibidem,idem,opcit,loccit}%
    \DeclareBibstringSetFormat{latin}{\mkbibemph{#1}}%
}

\AtEveryBibitem{%
    \clearfield{note}%
    \clearfield{pagetotal}%
}

\usepackage{siunitx}
\usepackage{mathtools}

\usepackage{graphicx}
\graphicspath{{./Figures/}{./Figures-Ext/}}

\usepackage{subcaption}
\usepackage{tikz}
\usetikzlibrary{shapes.geometric, positioning}

\usepackage{booktabs} 
\usepackage{tabularx} 

\usepackage{acro}
\acsetup{%
    use-id-as-short,
    single, 
    cite/display=first,
    uppercase/list=true}
\DeclareAcronym{QoS}{
    long=quality-of-service,
    single-style=long-short,
    tag=radio,
}
\DeclareAcronym{RSSI}{
    long=received signal strength index,
    tag=radio,
}
\DeclareAcronym{SINR}{
    long=signal-to-interference-plus-noise ratio,
    single-style=long-short,
    tag=radio,
}
\DeclareAcronym{RSRP}{
    long=reference signal received power,
    single-style=long-short,
    tag=radio,
}
\DeclareAcronym{RSRQ}{
    long=reference signal received quality,
    single-style=long-short,
    tag=radio,
}
\DeclareAcronym{CSI}{
    long=channel state information,
    single-style=short,
    tag=radio,
}
\DeclareAcronym{CQI}{
    long=channel quality indicator,
    single-style=long-short,
    tag=radio,
}
\DeclareAcronym{RI}{
    long=rank indicator,
    single-style=long-short,
    tag=radio,
}
\DeclareAcronym{PMI}{
    long=precoding matrix indicator,
    single-style=long-short,
    tag=radio,
}
\DeclareAcronym{EIRP}{
    long=effective isotropic radiated power,
    single-style=long-short,
    tag=radio,
}
\DeclareAcronym{MCS}{
    long=modulation and coding scheme,
    tag=radio,
}
\DeclareAcronym{QAM}{
    long=quadrature amplitude modulation,
    single-style=long-short,
    tag=radio,
}
\DeclareAcronym{HARQ}{
    long=hybrid automatic repeat request,
    single-style=long-short,
    tag=radio,
}
\DeclareAcronym{LOS}{
    long=line-of-sight,
    single-style=long-short,
    tag=radio,
}
\DeclareAcronym{NLOS}{
    long=non-line-of-sight,
    single-style=long-short,
    tag=radio,
}
\DeclareAcronym{TDD}{
    long=time-division duplexing,
    single-style=long-short,
    tag=radio,
}
\DeclareAcronym{IRS}{
    long=intelligent reflecting surface,
    tag=radio,
}
\DeclareAcronym{SBR}{
    long=shooting and bouncing rays,
    single-style=long-short,
    tag=radio,
}
\DeclareAcronym{DPM}{
    long=dominant path model,
    single-style=long-short,
    tag=radio,
}

\DeclareAcronym{5G}{
    long=fifth generation,
    list=fifth generation cellular network,
    single-style=long-short,
    cite=3GPP2025SystemArchitecture5GSystem,
    tag=radio,
}
\DeclareAcronym{EP5G}{
    long=Ericsson Private 5G,
    single-style=long-short,
    tag=radio,
}
\DeclareAcronym{5QI}{
    long=5G QoS identifier,
    single-style=long-short,
    tag=radio,
}
\DeclareAcronym{UE}{
    long=user equipment,
    single-style=long-short,
    tag=radio,
}
\DeclareAcronym{MIMO}{
    long=multiple-input multiple-output,
    single-style=long-short,
    tag=radio,
}
\DeclareAcronym{RAN}{
    long=radio access network,
    single-style=long-short,
    tag=radio,
}
\DeclareAcronym{PRB}{
    long=physical resource block,
    single-style=long-short,
    tag=radio,
}
\DeclareAcronym{eMBB}{
    long=enhanced mobile broadband,
    single-style=long-short,
    tag=radio,
}
\DeclareAcronym{URLLC}{
    long=ultra-reliable low-latency communication,
    single-style=long-short,
    tag=radio,
}
\DeclareAcronym{mMTC}{
    long=massive machine type communication,
    single-style=long-short,
    tag=radio,
}
\DeclareAcronym{SDN}{
    long=software defined networks,
    single-style=long-short,
    tag=networks,
}
\DeclareAcronym{NFV}{
    long=network function virtualization,
    single-style=long-short,
    tag=networks,
}
\DeclareAcronym{TCP}{
    long=transmission control protocol,
    single-style=short,
    tag=networks,
}
\DeclareAcronym{ICMP}{
    long=internet control message protocol,
    single-style=short,
    tag=networks,
}
\DeclareAcronym{VPN}{
    long=virtual private network,
    single-style=long-short,
    tag=networks,
}
\DeclareAcronym{V2X}{
    long=vehicular-to-everything,
    single-style=long-short,
    tag=radio,
}

\DeclareAcronym{3GPP}{
    long=The 3rd Generation Partnership Project,
    single-style=long-short,
    tag=radio,
}
\DeclareAcronym{5G-ACIA}{
    long=5G Alliance for Connected Industries and Automation,
    single-style=long-short,
    tag=meta,
}
\DeclareAcronym{5G-SMART}{
    long=5G for Smart Manufacturing,
    single-style=long-short,
    tag=meta,
}
\DeclareAcronym{5G NR}{
    long=New Radio,
    list=5G New Radio,
    single-style=short-long,
    cite=3GPP2023BaseStationRadioTransmissionReception,
    tag=radio,
}
\DeclareAcronym{LTE}{
    long=Long-Term Evolution,
    single-style=long-short,
    tag=radio,
}
\DeclareAcronym{ExPECA}{
    long=Experimental Platform for Edge Computing Applications,
    single-style=short-long,
    first-style=short-long,
    cite=Mostafavi2023ExPECAExperimentalPlatformTrustworthyEdge,
    tag=meta,
}
\DeclareAcronym{TECoSA}{
    long=The Center for Trustworthy Edge Computing Systems and Applications,
    single-style=long-short,
    first-style=long-short,
    tag=meta,
}
\DeclareAcronym{CRediT}{
    long=Contributor Roles Taxonomy,
    single-style=short,
    cite=Allen2019HowCanWeEnsureVisibility,
    tag=meta,
}

\DeclareAcronym{IoT}{
    long=internet-of-things,
    single-style=long-short,
    tag=networks,
}
\DeclareAcronym{CPS}{
    long=cyber-physical system,
}
\DeclareAcronym{HD}{
    long=high definition,
}
\DeclareAcronym{VR}{
    long=virtual reality,
}
\DeclareAcronym{AR}{
    long=augmented reality,
}
\DeclareAcronym{AGV}{
    long=automated guided vehicle,
}
\DeclareAcronym{AMR}{
    long=autonomous mobile robot,
}
\DeclareAcronym{AI}{
    long=artificial intelligence,
    single-style=long-short,
}

\DeclareAcronym{CAD}{
    long=computer-aided design,
    single-style=short,
}
\DeclareAcronym{LiDAR}{
    long=light detection and ranging,
    single-style=long-short,
}

\DeclareAcronym{PDDL}{
    long=The Planning Domain Definition Language,
    single-style=long-short,
    cite=McDermott1998PDDLPlanningDomainDefinitionLanguage,
    tag=planning,
}
\DeclareAcronym{STRIPS}{
    long=Stanford Research Institute Problem Solver,
    single-style=long-short,
    cite=Fikes1971STRIPSNewApproachApplicationTheorem,
    tag=planning,
}
\DeclareAcronym{POP}{
    long=partial-order planning,
    tag=planning,
}
\DeclareAcronym{HTN}{
    long=hierarchical task network,
    tag=planning,
}
\DeclareAcronym{HiPOP}{
    long=hierarchical partial-order planner,
    tag=planning,
}
\DeclareAcronym{SAT}{
    long=Boolean satisfiability,
    tag=planning,
}
\DeclareAcronym{SMT}{
    long=satisfiability modulo theories,
    tag=planning,
}

\DeclareAcronym{MDP}{
    long=Markov decision process,
    plural-form=Markov decision processes,
    tag=learning,
}
\DeclareAcronym{POMDP}{
    long=partially observable Markov decision process,
    plural-form=partially observable Markov decision processes,
    tag=learning,
}
\DeclareAcronym{RL}{
    long=reinforcement learning,
    tag=learning,
}
\DeclareAcronym{SMM}{
    long=shared mental model,
    tag=learning,
}
\DeclareAcronym{ANN}{
    long=artificial neural network,
    tag=learning,
}
\DeclareAcronym{GP}{
    long=Gaussian process,
    plural-form=Gaussian processes,
    tag=learning,
}
\DeclareAcronym{GPR}{
    long=Gaussian process regression,
    tag=learning,
}
\DeclareAcronym{RBF}{
    long=radial basis function,
    single-style=long-short,
    tag=learning,
}
\DeclareAcronym{M1}{
    short=Matérn,
    long=once-differentiable Matérn,
    single-style=long,
    tag=ignore,
}
\DeclareAcronym{RQ}{
    long=rational quadratic,
    single-style=long-short,
    tag=learning,
}

\DeclareAcronym{PDF}{
    long=probability density function,
    plural-form=probability density functions,
    single-style=long-short,
    tag=statistics,
}
\DeclareAcronym{RMSE}{
    long=root mean squared error,
    single-style=long-short,
    tag=statistics,
}
\DeclareAcronym{MAE}{
    long=mean absolute error,
    single-style=long-short,
    tag=statistics,
}
\DeclareAcronym{MAD}{
    long=median absolute deviation,
    single-style=long-short,
    tag=statistics,
}
\DeclareAcronym{WAPE}{
    long=weighted absolute percentage error,
    single-style=long-short,
    tag=statistics,
}
\DeclareAcronym{SD}{
    long=standard deviation,
    single-style=long,
    tag=statistics,
}
\DeclareAcronym{KNN}{
    long=$k$-nearest neighbor,
    plural-form=$k$-nearest neighbors,
    single-style=long,
    tag=statistics,
}

\DeclareAcronym{MPC}{
    long=model-predictive control,
}
\DeclareAcronym{DP}{
    long=dynamic programming,
}
\DeclareAcronym{MILP}{
    long=mixed-integer linear program,
}
\DeclareAcronym{RRT}{
    long=rapidly-exploring random trees,
    plural=rapidly-exploring random trees,
}
\DeclareAcronym{RRT*}{
    long=asymptotically optimal rapidly-exploring random trees,
    plural=asymptotically optimal rapidly-exploring random trees,
    cite=Karaman2011SamplingbasedAlgorithmsOptimalMotionPlanning
}
\DeclareAcronym{POI}{
    long=point-of-interest,
    plural-form=points-of-interest,
}
\DeclareAcronym{MAPF}{
    long=multi-agent pathfinding,
}
\DeclareAcronym{APF}{
    long=artificial potential field,
}

\DeclareAcronym{CTL}{
    long=computation tree logic,
    single-style=long-short,
}
\DeclareAcronym{LTL}{
    long=linear temporal logic,
    single-style=long-short,
}
\DeclareAcronym{STL}{
    long=signal temporal logic,
    single-style=long-short,
    cite=Maler2004MonitoringTemporalPropertiesContinuousSignals
}
\DeclareAcronym{SpaTeL}{
    long=spatio-temporal logic,
    single-style=long-short,
    cite=Haghighi2015SpaTeLNovelSpatialtemporalLogicIts
}
\DeclareAcronym{STREL}{
    long=spatio-temporal reach and escape logic,
    single-style=long-short,
    cite=Bartocci2017MonitoringMobileSpatiallyDistributedCyberphysical
}

\usepackage[hidelinks]{hyperref}
\usepackage[capitalise,nameinlink]{cleveref}

\crefname{equation}{}{}
\Crefname{equation}{Equation}{Equations}
\crefname{figure}{Fig.}{Figs.}
\Crefname{figure}{Fig.}{Figs.}

\usepackage{enumitem}

\usepackage{xcolor}

\makeatletter
\def\convertto#1#2{\strip@pt\dimexpr #2*65536/\number\dimexpr 1#1}
\makeatother

\newcommand{\email}[1]{\href{mailto:#1}{#1}}

\begin{document}

\title{%
    Why Channel-Centric Models are not Enough to Predict End-to-End Performance in Private 5G:\\A Measurement Campaign and Case Study%
}

\author{%
    Nils~Jörgensen \\
    School of Industrial Engineering and Management\\
    KTH Royal Institute of Technology\\
    Stockholm, Sweden\\
    Email: \email{nilsjor@kth.se}
    \thanks{%
        Contributor role (\acs{CRediT}) statement: 
        Methodology, Investigation, and Data Curation: Awad AlNasrallah and Hoahui Meng; 
        Supervision and Review \& Editing: Prof. Fredrik Asplund and Dr. Ajay Kattepur; 
        Resources: Stefan Rönngren.
        This research has received support from \href{https://www.tecosa.center.kth.se}{\acs{TECoSA}}.
        This work has been submitted to the IEEE for possible publication. 
        Copyright may be transferred without notice, after which this version may no longer be accessible.
    }
}

\maketitle

\begin{abstract}
Communication-aware robot planning requires accurate predictions of wireless network performance.
Current approaches rely on channel-level metrics such as received signal strength and signal-to-noise ratio, assuming these translate reliably into end-to-end throughput.
We challenge this assumption through a measurement campaign in a private 5G industrial environment.
We evaluate throughput predictions from a commercial ray-tracing simulator as well as data-driven Gaussian process regression models against measurements collected using a mobile robot.
The study uses off-the-shelf user equipment in an underground, radio-shielded facility with detailed 3D modeling, representing a best-case scenario for prediction accuracy.
The ray-tracing simulator captures the spatial structure of indoor propagation and predicts channel-level metrics with reasonable fidelity.
However, it systematically over-predicts throughput, even in line-of-sight regions.
The dominant error source is shown to be over-estimation of sustainable MIMO spatial layers: the simulator assumes near-uniform four-layer transmission while measurements reveal substantial adaptation between one and three layers.
This mismatch inflates predicted throughput even when channel metrics appear accurate.
In contrast, a Gaussian process model with a rational quadratic kernel achieves approximately two-thirds reduction in prediction error with near-zero bias by learning end-to-end throughput directly from measurements.
These findings demonstrate that favorable channel conditions do not guarantee high throughput; communication-aware planners relying solely on channel-centric predictions risk overly optimistic trajectories that violate reliability requirements.
Accurate throughput prediction for 5G systems requires either extensive calibration of link-layer models or data-driven approaches that capture real system behavior.
\end{abstract}

\begin{IEEEkeywords}
5G mobile communication, 
Throughput, 
MIMO, 
Channel models, 
Ray tracing, 
Gaussian processes
\end{IEEEkeywords}


\begin{table}[h]
    \caption{List of abbreviations used in this paper.}
    \label{tab:ListOfAbbreviations}
    \centering
    \small
    \printacronyms[exclude={ignore,meta}, template=tabular, heading=none]
    \vspace{-2em}
\end{table}

\section{Introduction}
\label{sec:Introduction}

Industrial automation is increasingly driven by mobile and collaborative robots that rely on reliable wireless connectivity for tasks such as supervision, teleoperation, and cloud/edge offloading.
\Ac{5G} systems are being positioned as a key enabler of this trend, promising high data rates, low latency, and \ac{QoS} guarantees tailored to industrial use cases~\cite{
    5G-ACIA20215GQoSIndustrialAutomation,
    Mahmood2022Factory5GReviewIndustrycentricFeatures,
    Wollschlaeger2017FutureIndustrialCommunicationAutomationNetworks}.
In factory environments, private \ac{5G} networks are being deployed to support \acp{AMR} and \acp{AGV}, often operating in cluttered, metal-rich environments with strong multipath and blockage effects~%
\cite{
    5G-ACIA2020Key5GUseCasesRequirements,
    5G-SMART2021ReportDevelopment5GUseCases,
    Chen2021WirelessNetworkedMultirobotSystemsSmart}.
Consequently, the ability to accurately predict radio performance is crucial for both network design and the safe, reliable operation of mobile robots.

\begin{table*}[ht]
    \centering
    \normalsize 
    \caption{Positioning of this study relative to prior work across key methodological dimensions.}
    \label{tab:IntroRelatedWork}
    \begin{tabular}{llcccc}
    \toprule
    \textbf{Reference} & \textbf{Focus area} & \textbf{Indoor} & \textbf{Throughput} & \textbf{Measured} & \textbf{Predictive} \\ \midrule
    \textcite{Sheikh2021BlockageRayTracingPropagationModel} & Ray tracing simulation & \checkmark & \checkmark &   & \checkmark \\
    \textcite{3GPP2022StudyChannelModelFrequencies05} & Stochastic channel model & \checkmark &   &   & \checkmark \\
    \textcite{Unger2025ImpactTerrainSamplingDensity5G} & Propagation model comparison &   &   &   & \checkmark \\
    \textcite{Okano2020FieldExperiments28GHzBand} & 5G field trial & \checkmark & \checkmark & \checkmark &   \\
    \textcite{5G-SMART2022Report5GRadioDeployabilityFactory} & Industrial 5G deployment & \checkmark &   & \checkmark &   \\
    \textcite{Cantero2023SystemlevelPerformanceEvaluation5GUse} & System-level simulation & \checkmark & \checkmark &   & \checkmark \\
    \textcite{Vijayan20235GWirelessChannelCharacterizationIndoor} & InF channel validation & \checkmark &   & \checkmark & \checkmark \\
    \textcite{Polak2024MeasurementAnalysis4G5GMobile} & Industrial coverage & \checkmark & \checkmark & \checkmark &   \\
    \textcite{Hernangomez2024AIenabledConnectedIndustryAGVCommunication} & AGV measurement dataset & \checkmark &   & \checkmark &   \\
    \textcite{Tarneberg2020IntelligentIndustry405GNetworks} & Outdoors QoE measurement &   & \checkmark & \checkmark &   \\
    \textcite{Santra2021ExperimentalValidationDeterministicRadioPropagation} & Propagation model validation &   &   & \checkmark & \checkmark \\
    \textcite{Coll-Perales2022EndtoendV2XLatencyModelingAnalysis} & V2X latency modeling &   &   &   & \checkmark \\
    \textcite{Lazar2023RealtimeDataMeasurementMethodologyEvaluate} & 5G KPI measurement &   & \checkmark & \checkmark &   \\
    \textcite{Costa2023ThroughputSlopesPrediction5GNetworks} & Throughput prediction &   & \checkmark &   & \checkmark \\ \midrule
    This work & Industrial 5G robotics & \checkmark & \checkmark & \checkmark & \checkmark \\ \bottomrule
\end{tabular}
\vspace{-1em}
\end{table*}

This need has given rise to \emph{communication-aware} motion planning, an emerging paradigm in which predicted channel quality informs robot trajectories and task allocation~\cite{Muralidharan2021CommunicationawareRoboticsExploitingMotionCommunication}.
Representative approaches include 
    resilient communication-aware motion planners~\cite{Caccamo2017RCAMPResilientCommunicationawareMotionPlanner}, 
    joint motion--communication co-optimization~\cite{Ali2019MotioncommunicationCooptimizationCooperativeLoadTransfer}, 
    decentralized multi-robot coordination with inter-robot relay links~\cite{Liu2020DistributedCommunicationawareMotionPlanningNetworked},
    and radio-map-based path planning augmented with intelligent reflecting surfaces~\cite{Mu2021IntelligentReflectingSurfaceEnhancedIndoor}.
These methods typically employ parametric or learned models of channel quantities---such as received signal strength, path loss, or \ac{SINR}---and treat them as cost terms or constraints in the planning problem, implicitly assuming that favorable channel conditions translate into adequate end-to-end performance.

Yet for autonomous robots, the metrics of ultimate interest are often end-to-end throughput and latency along a trajectory, rather than \ac{SINR} or \ac{RSRP} in isolation.
While latency is central to many time-critical robotic applications, it can largely be controlled by preventing throughput from saturating; thus, accurate throughput prediction becomes a key enabler for latency-aware planning.
In operational \ac{5G} systems, however, throughput depends not only on large-scale propagation but also on link adaptation, spatial multiplexing, precoding, scheduler behavior, and protocol overhead.
Whether channel-level metrics reliably predict throughput is therefore an empirical question---one that has received surprisingly little attention.

This is not for lack of modeling tools.
A substantial body of work has investigated \ac{5G} channel measurements and modeling across diverse scenarios---see~\cite{
    Sarkar2003SurveyVariousPropagationModelsMobile,
    Wang2018Survey5GChannelMeasurementsModels} 
for a comprehensive overview---leading to the development of standardized empirical stochastic models, such as 3GPP TR~38.901~\cite{
    3GPP2022StudyChannelModelFrequencies05}.
In parallel, site-specific deterministic ray-tracing and hybrid models have been developed and validated for indoor factories and other complex environments~\cite{
    Santra2021ExperimentalValidationDeterministicRadioPropagation,
    Sheikh2021BlockageRayTracingPropagationModel,
    Qin2021EfficientModelingRadioWavePropagation,
    Vijayan20235GWirelessChannelCharacterizationIndoor}.
Complementing these physics-based approaches, data-driven methods---including machine-learning radio maps and \ac{GP} models---have been proposed to predict path loss and related quantities such as coverage, signal strength, and theoretical spectral efficiency directly from measurements~\cite{
    Bi2019EngineeringRadioMapsWirelessResource,
    Olukanni2023GaussianProcessRegressionModelPredict,
    Kallehauge2024SpatialPredictionWirelessChannelStatistics}.
These tools are well established for predicting channel-level quantities; whether they---or any model---can accurately predict end-to-end throughput remains an open question.

\subsection{Related Work}
Validation of radio propagation models through measurement studies remains an active area of research.
A comprehensive survey by \textcite{Wang2018Survey5GChannelMeasurementsModels} reviews the state of \ac{5G} channel measurements and models, covering diverse scenarios including massive \ac{MIMO}, millimeter-wave communications, and mobility.
That survey identifies proper \ac{MIMO} modeling as essential for \ac{5G} channels, emphasizing channel-level properties such as spherical wavefront and array non-stationarity.
Crucially, however, the survey---and the body of work it reviews---focuses predominantly on channel-level quantities: path loss, delay spread, angular spread, and related propagation characteristics.

This channel-centric focus persists in subsequent work: comparisons between 3GPP-type stochastic models and ray-tracing in indoor factory-like environments typically report \ac{RSRP}, \ac{SINR}, or path loss as the primary indicators of model fidelity~\cite{
    Sheikh2021BlockageRayTracingPropagationModel,
    3GPP2022StudyChannelModelFrequencies05,
    Unger2025ImpactTerrainSamplingDensity5G}.
A growing number of measurement campaigns investigate \ac{5G} coverage, latency, and reliability for industrial scenarios, with a focus on system-wide performance indicators~\cite{
    Okano2020FieldExperiments28GHzBand,
    5G-SMART2022Report5GRadioDeployabilityFactory,
    Cantero2023SystemlevelPerformanceEvaluation5GUse,
    Vijayan20235GWirelessChannelCharacterizationIndoor,
    Polak2024MeasurementAnalysis4G5GMobile,
    Hernangomez2024AIenabledConnectedIndustryAGVCommunication}.
Other studies emphasize outdoor or vehicular scenarios~\cite{
    Tarneberg2020IntelligentIndustry405GNetworks,
    Santra2021ExperimentalValidationDeterministicRadioPropagation,
    Coll-Perales2022EndtoendV2XLatencyModelingAnalysis,
    Lazar2023RealtimeDataMeasurementMethodologyEvaluate,
    Costa2023ThroughputSlopesPrediction5GNetworks}.
\Cref{tab:IntroRelatedWork} positions this study relative to prior work across four methodological dimensions:
(1)~whether the study targets \emph{indoor} industrial or factory-like environments;
(2)~whether \emph{throughput} is evaluated as a key performance metric;
(3)~whether results are based on real-world \emph{measurements} rather than pure simulation;
and (4)~whether the work develops or validates \emph{predictive} models.
Strikingly, no prior work evaluates a prediction model's ability to provide actionable throughput estimates---whether predicted directly or derived from channel-level metrics such as \ac{SINR}.
The assumption that favorable channel conditions guarantee adequate throughput thus remains empirically untested, despite its centrality to communication-aware robot planning.

\subsection{Purpose and Contributions}
\label{sec:IntroductionPurpose}
This paper addresses this gap through a single-site case study of downlink \ac{5G} throughput in a realistic industrial-like environment.
We consider an underground, radio-shielded hall instrumented with a private \ac{5G} network and a mobile robot, and construct site-wide throughput maps combining physics-based simulation and on-site measurements.
A commercial ray-tracing simulation tool, configured with a 3GPP-compliant industrial channel model, is used to generate deterministic predictions, while \ac{GP} models are trained on measurement data to provide data-driven throughput estimates.
We then compare these approaches using a unified analysis framework and discuss their implications for communication-aware robot planning.
%
Specifically, we address the following research questions:
\begin{description}
    \item[RQ1:] How accurately do physics-based ray-tracing simulators predict \ac{5G} throughput, and what are the significant sources of prediction error?
    \item[RQ2:] How do data-driven \ac{GPR} models compare to physics-based simulators for throughput prediction, and what factors influence their predictive performance the most?
\end{description}

Our results reveal that the physics-based ray-tracing simulator exhibits systematic over-prediction of throughput, driven primarily by over-estimation of sustainable \ac{MIMO} spatial layers rather than inaccurate channel-level metrics.
In contrast, using \ac{GPR} achieves approximately two-thirds reduction in prediction error with near-zero bias, by learning end-to-end throughput behavior directly from measurements without explicit rank modeling.
These findings indicate that communication-aware planners relying on channel-centric predictions may produce \emph{overly optimistic} trajectory plans in regions with favorable \ac{SINR} conditions.

The main contributions of this work are as follows:
\begin{itemize}
    \item We conduct a robotic measurement campaign in a private indoor \ac{5G} deployment, collecting spatially dense downlink throughput, latency, and link-layer measurements in a complex, industrial-like environment, and make the dataset publicly available~\cite{AlNasrallah2025R1EP5GChannelMeasurement}.
    \item We configure a state-of-the-art ray-tracing-based radio network planning tool for the same site, combining detailed 3D environment modeling, 3GPP-compliant indoor-factory channel settings, and parametric link modeling to obtain simulated throughput predictions.
    \item We develop data-driven throughput models using \ac{GPR}, and train them on the collected measurements.
    \item We propose and apply a comparative analysis framework that quantifies the bias, accuracy, and variance of throughput predictions for both parametric and data-driven models.
    \item We highlight conditions under which channel-level metrics fail to predict throughput reliably, and discuss how these findings affect joint communication--motion planning for mobile robots.
\end{itemize}

\subsection{Structure of this Paper}
The remainder of this paper is organized as follows.
\Cref{sec:Background} reviews industrial \ac{5G}, radio-channel modeling, and communication-aware motion planning.
\Cref{sec:Methods} presents the ray-tracing simulation, robotic measurement campaign, data-driven throughput modeling, and comparative analysis framework.
\Cref{sec:Results} presents the results with some analysis.
\Cref{sec:Discussion} discusses the results and puts forth relevant conclusions to the literature.
\Cref{sec:Conclusion} summarizes the conclusions of the study and outlines opportunities for future work.

\section{Background}
\label{sec:Background}

This study centers on \emph{throughput}---the volume of data successfully delivered over a wireless channel per unit time (e.g., Mbps)---as the primary performance metric.
Throughput directly reflects the end-user experience and is sensitive to network resource allocation, particularly the dynamic assignment of \acp{PRB} at the \ac{RAN} level.
Consequently, it serves as a natural objective for joint communication--motion planning, where robot trajectories may be optimized against predicted data-rate maps.

\subsection{Radio Channel Modeling}
Radio-channel models broadly fall into two categories.
\emph{Empirical} (statistical) models, such as those specified in 3GPP TR~38.901~\cite{3GPP2022StudyChannelModelFrequencies05}, offer computational efficiency and require no detailed knowledge of the physical environment, but must be calibrated from measurement data.
\emph{Deterministic} models, such as ray-tracing, instead derive channel characteristics from explicit three-dimensional geometry and material properties, trading higher computational cost for site-specific accuracy without the need for prior measurements.
For one part of this study, we employ a deterministic ray-tracing approach configured with 3GPP-compliant indoor-factory parameters.

Both approaches are fundamentally \emph{channel-centric}: they predict quantities such as path loss or \ac{SINR}, from which throughput is inferred through additional link-layer modeling.
An alternative is to learn throughput maps directly from measurement data using data-driven regression techniques, bypassing the intermediate channel abstraction altogether.

\subsection{Data-driven Modeling using \acs{GPR}}
\Ac{GPR} provides a probabilistic, non-parametric framework for modeling continuous spatial fields from data.
Unlike physics-based or purely parametric models, \ac{GPR} does not rely on explicit assumptions about the propagation environment or system behavior.
Given training data, the model produces smooth spatial estimates together with predictive variances that quantify uncertainty---for example, in sparsely sampled or unobserved regions.
These properties make \ac{GPR} applicable to scenarios where multipath effects and adaptive link mechanisms are difficult to model deterministically.

In this study, \ac{GP} models are trained using spatial measurement data consisting of a 2D-coordinate, a mean throughput, and the associated variability metric derived from repeated measurements (for implementation details, see \cref{sec:MethodsDataDriven}).
The training process fits the model to this dataset by adjusting its hyperparameters---such as signal variance, length scale, and noise variance---to maximize the marginal log-likelihood of the observations.
Once trained, the \ac{GP} model provides predictive means and variances for any query location, enabling extrapolation across unmeasured regions together with a calibrated estimate of prediction confidence.

\subsection{Kernel Functions}
\label{sec:BackgroundKernelFunctions}
Rather than specifying a fixed functional form, \ac{GP} models define a prior distribution over functions
\begin{equation}
    \label{eq:BackgroundGprPrior}
    f(x) \sim \mathcal{GP}\left(m(x), k(x,x')\right),
\end{equation}
where $m(x)$ is the mean function and $k(x,x')$ is the covariance kernel \cite[Sec. 2.2]{Rasmussen2008GaussianProcessesMachineLearning}.
The choice of kernel directly affects the covariance structure and thus the smoothness, length-scale behavior, and extrapolation properties of the trained models.

In this study, each evaluated kernel is formed as the product of a constant kernel and a stationary base kernel, with an additive white-noise kernel to capture measurement uncertainty, preventing the model from interpolating exactly through noisy observations and enabling smooth generalization between training points.
This composite-kernel construction is inspired by \cite[Chp. 2]{Duvenaud2014AutomaticModelConstructionGaussianProcesses}.
The general form can be written as
\begin{equation}
    \label{eq:BackgroundCompositeKernel}
    k(x, x') = \sigma_f^2 \cdot k_\text{base}(x, x') + \sigma_n^2 \cdot \delta(x, x'),
\end{equation}
where $\sigma_f^2$ is the signal variance, scaling the base kernel $k_\text{base}(x, x')$; $\sigma_n^2$ is the white-noise level and $\delta(x, x')$ the Kronecker delta function.
Below are the definitions of the three base kernels used in this study. 
They are common choices, as seen in 
\cite{Xie2020AnalyzingMachineLearningModelsGaussian, 
Salahshoori2024MachineLearningpoweredEstimationMalachiteGreen,
Rossmann2025GenericSingleobjectiveMachineDesignFramework}.

\subsubsection*{\Acf{RBF}}
The \acs{RBF} (or ``squared exponential'') kernel can be written as
\begin{equation}
    \label{eq:BackgroundRbfKernel}
    k_\text{RBF}(x, x') = \exp\left(-\frac{\|x - x'\|^2}{2\ell^2}\right),
\end{equation}
and measures the similarity between two points in space:
\begin{itemize}
    \item if $x$ and $x'$ are very close, the kernel value is near 1;
    \item if they are far apart, the kernel value drops toward 0.
\end{itemize}
The characteristic length scale $\ell$ is optimized during the training process, and determines how quickly the correlation between points decays. 
This kernel assumes that the underlying function is \emph{infinitely differentiable}, making it useful for modeling smooth and continuous phenomena.

\subsubsection*{\Acl{M1}}
The Matérn kernel is a generalization of the \ac{RBF} kernel, controlled by a fixed smoothness parameter $\nu > 0$.
Choosing $\nu = 1.5$ yields a \emph{once-differentiable} covariance function, balancing moderately rough spatial variation with overall regularity.
The resulting kernel is defined as
\begin{equation}
    \label{eq:BackgroundMaternKernel}
    k_{\text{M1}}(x, x') = \left(1 + \frac{\sqrt{3}\,\|x - x'\|}{\ell}\right) \cdot \exp\left(-\frac{\sqrt{3}\,\|x - x'\|}{\ell}\right).
\end{equation}
%

\subsubsection*{\Acf{RQ}}
The \acs{RQ} is another smooth but flexible generalization of the \acs{RBF} kernel that better captures multi-scale structure in data.
It can be written as
\begin{equation}
    \label{eq:BackgroundRqKernel}
    k_\text{RQ}(x, x') = \left(1 + \frac{\|x - x'\|^2}{2\,\alpha\,\ell^2}\right)^{-\alpha}.
\end{equation}
This kernel can be interpreted as an \emph{infinite mixture} of \acs{RBF} kernels, integrated over an inverse gamma distribution on the squared length scale.
The scale mixture parameter $\alpha>0$ controls how wide that distribution of length scales is:
\begin{itemize}
    \item small $\alpha$: broad mixture, allows both short- and long-scale correlations, captures multi-scale structure;
    \item large $\alpha$: narrow mixture, similar to \acs{RBF} kernel.
\end{itemize}
This parameter $\alpha$ is co-optimized with $\ell$ during training.

\section{Methods}
\label{sec:Methods}

As outlined in \cref{sec:IntroductionPurpose}, we conduct a single-site case study to compare a physics-based radio-channel simulator with a data-driven model for predicting \ac{5G} downlink throughput in a realistic industrial environment.

\begin{figure}[htb]
    \centering
    \includegraphics[width=\columnwidth]{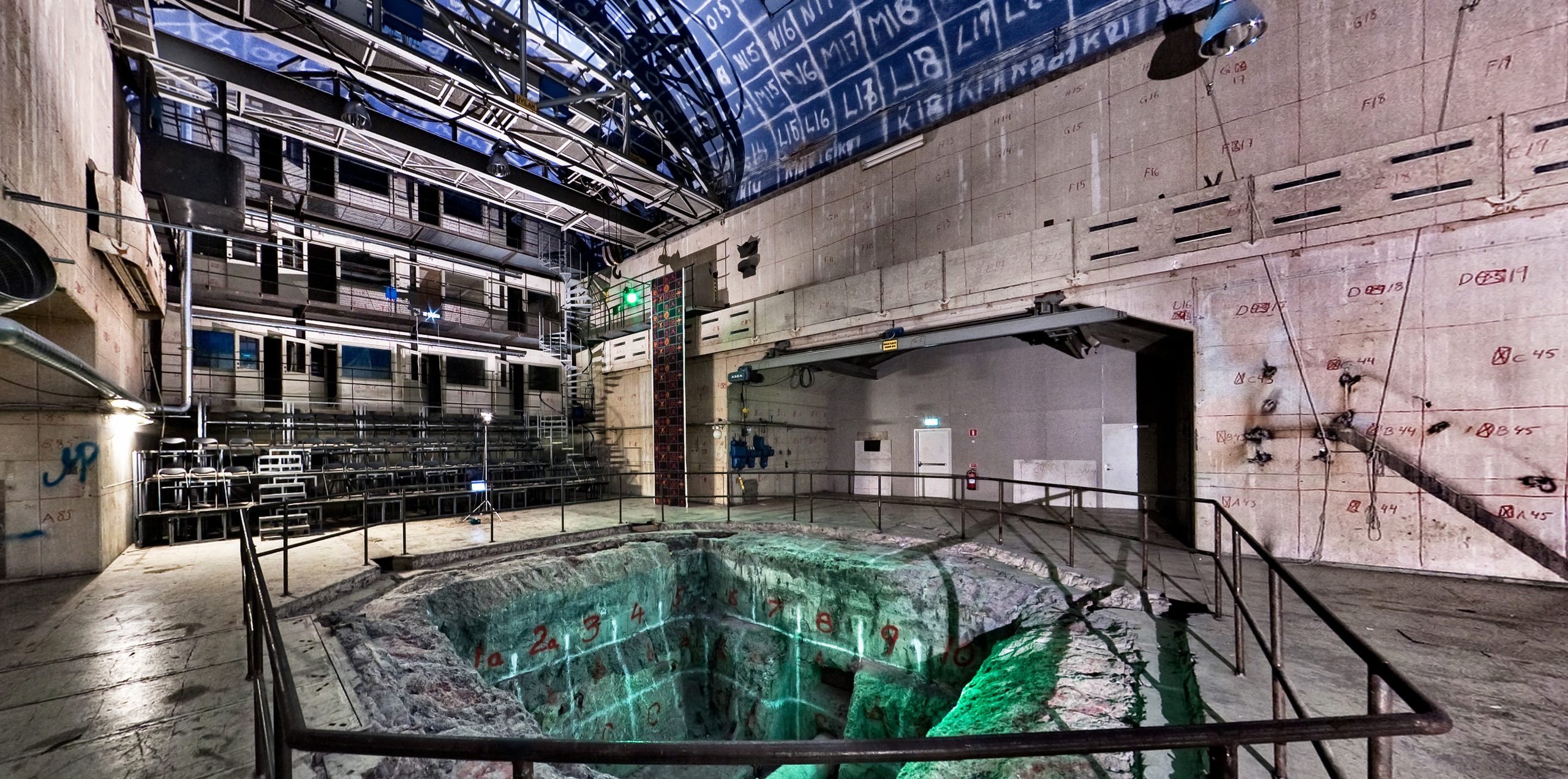}
    \caption{%
        Photograph of the testbed site; an underground, radio-shielded facility used for edge- and cloud-computing research. 
        Photo © \citeyear{JannLipka2008Reaktor1KTHEntranceView360} \citeauthor{JannLipka2008Reaktor1KTHEntranceView360}, used with permission. 
        Source: 360Cities \cite{JannLipka2008Reaktor1KTHEntranceView360}.}
    \label{fig:MethodsR1Photo}
\end{figure}

The study site is the KTH Reactor Hall, an underground facility that once housed Sweden's first nuclear reactor \cite{KTH2023HistoryKTHReactorHall}.
Today, it features an \ac{EP5G} network deployed as part of \ac{ExPECA}, a state-of-the-art research testbed providing high-performance edge computing and a range of networking capabilities.
\Cref{fig:MethodsR1Photo} shows the testbed site and the workspace in which the robot navigated during the measurement campaign.

The \ac{ExPECA} testbed was selected because it 
(1) integrates a private \ac{5G} core and radio equipment in a controlled environment, 
(2) provides a radio-shielded, large-volume indoor space that eliminates external interference, and
(3) exposes container orchestration and isolated virtual networks, simplifying reproducible experiment deployment.

The Reactor Hall contains multiple radio dots, serving as base stations, at different elevations.
This study intentionally uses a \emph{single} high-mounted radio dot (approximately \qty{8}{\meter} elevation; bottom-left position in the plan view; see \cref{fig:MethodsRadioDots}) to create a more challenging coverage scenario and to stress-test prediction methods under complex indoor propagation.

The computational electromagnetics suite Altair Feko \cite{Hoppe2017WavePropagationRadioNetworkPlanning} was selected for deterministic radio propagation modeling because of its 3D CAD integration, native \ac{5G} air-interface support, versatile ray-tracing engine, and parametric antenna modeling utilities.

\begin{figure}[htb]
    \centering
\newlength{\tempdimen}
\setlength{\tempdimen}{0.95\columnwidth / 14}  
\pgfmathsetmacro{\scalefactor}{\the\tempdimen / 1cm}  

\begin{tikzpicture}[xscale=\scalefactor, yscale=0.8*\scalefactor]
    \tikzset{
        every node/.style={
            draw, 
            fill=blue!10, 
            rounded corners=\scalefactor cm * 0.2, 
            align=center,
            inner sep=0pt, 
            minimum width=\scalefactor cm * 2, 
            text width=\scalefactor cm * 2*0.9, 
            minimum height=\scalefactor cm * 2*0.618, 
            font=\sffamily\fontsize{5}{5}\selectfont  
        }
    }
    
    \node(model) at (0,0)
      {\hyperref[sec:MethodsEnvironmentModeling]{\nameref*{sec:MethodsEnvironmentModeling}}};
    \node[font=\sffamily\fontsize{5}{4.5}\selectfont](collect) at (1,-2)
      {\hyperref[sec:MethodsRadioAccess]{\nameref*{sec:MethodsRadioAccess}}};
    \node(simulate) at (4,-1)
      {\hyperref[sec:MethodsPropagationSimulation]{\nameref*{sec:MethodsPropagationSimulation}}};
    \node(measure) at (5,-3)
      {\hyperref[sec:MethodsRoboticCampaign]{\nameref*{sec:MethodsRoboticCampaign}}};
    \node(compare) at (9,0)
      {\hyperref[sec:MethodsComparativeFramework]{\nameref*{sec:MethodsComparativeFramework}}};
    \node(train) at (8,-2)
      {\hyperref[sec:MethodsDataDriven]{\nameref*{sec:MethodsDataDriven}}};
    \node(analyze) at (12,-1)
      [fill=green!10] {\hyperref[sec:Results]{Results Visualization and Analysis}};

    \draw[-latex] (model.east) -- ++(1.0,0) |- ([yshift=0.25cm]simulate.west);
    \draw[-latex] (collect.east) -- ++(0.5,0) |- ([yshift=-0.25cm]simulate.west);
    \draw[-latex] (collect.east) ++(0.5,0) |- (measure.west); 
    \draw[-latex] (simulate.east) -- ++(1.0,0) |- ([yshift=0.25cm]compare.west);
    \draw[-latex] (measure.east) -- ++(0.5,0) |- ([yshift=-0.25cm]compare.west);
    \draw[-latex] (measure.east) ++(0.5,1) |- (train.west); 
    \draw[-latex] (compare.east) -- ++(0.5,0) |- ([yshift=0.25cm]analyze.west);
    \draw[-latex] (train.east) -- ++(1.0,0) |- ([yshift=-0.25cm]analyze.west);
\end{tikzpicture}
    \caption{Overview of the experimental and modeling workflow.}
    \label{fig:MethodsWorkflow}
\end{figure}
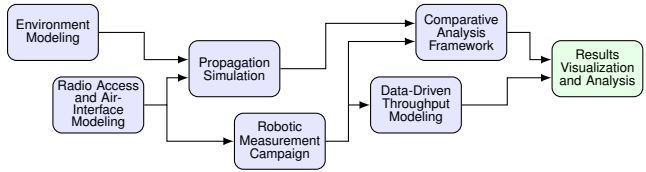

\Cref{fig:MethodsWorkflow} illustrates the overall workflow of the study, where each stage corresponds to a subsequent section below.
It outlines the sequential stages of the study---from modeling and radio configuration to simulations, robotic data collection, and data-driven modeling---forming a unified framework for comparing model predictions with real-world measurements.
The final step (shown in green) corresponds to \cref{sec:Results}, where we visualize and compare predictions against ground-truth measurements.

\subsection{Environment Modeling}
\label{sec:MethodsEnvironmentModeling}
A detailed 3D model of the Reactor Hall was available from previous work by Adapt AB \cite{Felldin2022DagPaAdaptArSig}.
The model was created using a BLK360 3D scanner to capture the unique architecture of the facility with high precision. 
This specialized equipment generated point clouds with approximately \qty{5}{\milli\meter} precision, which were then imported into computer design software for manual refinement to accurately represent architectural details and major fixtures.
To reduce computational load while preserving key geometry, we simplified the original mesh for this study from approximately \qty{1.1}{M} polygons to just under \num{100000} polygons prior to simulation.
Pre-defined material properties were assigned to surfaces, and the resulting environment database was imported for propagation simulation.

\begin{figure}[ht]
    \centering
    \includegraphics[width=1\columnwidth]{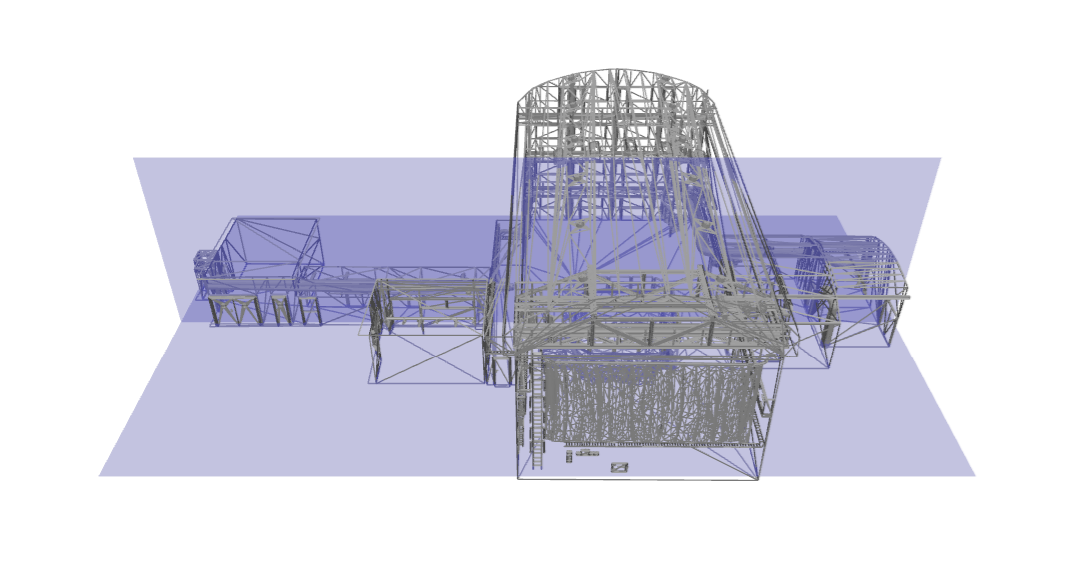}
    \caption{3D model of the KTH Reactor Hall imported into Altair Feko for radio propagation simulation.}
    \label{fig:MethodsWinprop3dModel}
\end{figure}

\Cref{fig:MethodsWinprop3dModel} illustrates the simplified 3D geometry used for ray tracing, which preserved major reflectors and occluders while removing small-scale details that minimally affect large-scale propagation patterns.

\subsection{Radio Access and Air-Interface Modeling}
\label{sec:MethodsRadioAccess}
Building on the environment model, we configured the air-interface parameters and antenna placement to match the \ac{ExPECA} deployment.
Key configuration parameters include:
\begin{enumerate}
    \item \textbf{Topology:} Four ceiling-mounted radio dots, which can be individually enabled or disabled.
    \item \textbf{\acs{MIMO} capabilities:} Each dot supports $4\times 4$ \acs{MIMO} downlink and $2\times 2$ \acs{MIMO} uplink.
    \item \textbf{Carrier and numerology:} Band~n78, center frequency $f_{\text{c}}=\qty{3760}{\mega\hertz}$, channel bandwidth \qty{80}{\mega\hertz}, and \acs{3GPP} numerology $\mu=1$ (i.e., \qty{30}{\kilo\hertz} subcarrier spacing).
    \item \textbf{\acs{TDD} frame pattern:} Pattern~2, symbol sequence D-D-D-S-U, with Sub-pattern~2 (10:2:2); see \cite{3GPP2023PhysicalLayerProceduresControl} for details.
\end{enumerate}

The modem used in this study reports only downlink metrics; consequently, the study is confined to downlink throughput measurements.
However, because the uplink/downlink slot allocations are fixed, the effect of this restriction on the comparative analysis is expected to be limited.

\begin{figure}[ht]
    \centering
    \includegraphics[width=1\columnwidth]{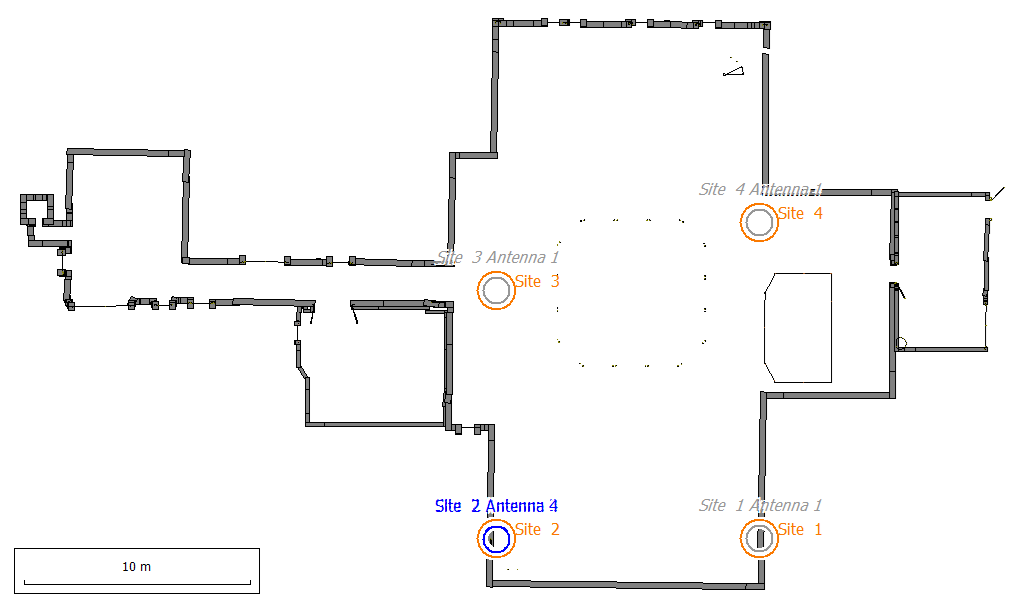}
    \caption{Plan view of the Reactor Hall showing the placement of all radio dots in the environment. Only the highlighted dot (bottom-left) was enabled for the measurement campaign.}
    \label{fig:MethodsRadioDots}
\end{figure}

\Cref{fig:MethodsRadioDots} shows the plan view with the location of the single active radio dot used for measurements.
The dashed line in the center of the hall marks the "pit"---a recessed, non-traversable area where the nuclear reactor once operated, now enclosed by a protective railing. 

\subsection{Propagation Simulation}
\label{sec:MethodsPropagationSimulation}
With the physical environment and air-interface configured, we employed the \ac{SBR} method in Altair Feko for propagation prediction.
\ac{SBR} was selected over other methods for its ability to capture reflections, diffractions, transmissions, and scattering in complex indoor environments while balancing computational cost against simulation fidelity.

The simulation was executed on a \qty{0.5}{\meter} horizontal grid map covering the accessible floor area.
This generated spatial predictions of coverage and derived link-performance indicators, as well as predicted maximum throughput
Post-processing exported these predictions as heatmaps for subsequent comparison against measurements, with \ac{MCS} indices inferred via lookup from throughput fields for completeness of analysis.

\subsection{Robotic Measurement Campaign}
\label{sec:MethodsRoboticCampaign}
To validate the simulation predictions and provide training data for the data-driven models, we conducted a comprehensive measurement campaign using a modified TurtleBot3 equipped with a Raspberry Pi 4 and a Waveshare 5G hat featuring a Quectel RM500Q-GL modem.

For experimental control, we configured isolated virtual networks on the \ac{ExPECA} testbed to ensure measurement traffic traversed the \ac{5G} core network via a dedicated path with minimal hops between the robot and edge server.
Husarnet \cite{Husarnet2025HusarnetOperateEdgeLatency}, an IPv6-based overlay \ac{VPN}, was used exclusively for control and management traffic, keeping measurement flows on the native \ac{EP5G} data path to avoid introducing artificial performance artifacts resulting from tunneling.

To minimize confounding from small-scale fading and mobility, we employed a static sampling protocol.
The robot advanced in discrete \qty{0.5}{\meter} steps, stopping at each waypoint to collect:
\begin{itemize}
    \item \textbf{Physical/link-layer metrics:} Radio channel metrics, \ac{MCS}, and effective data layers using modem AT commands at \qty{3}{\hertz}; see \cref{tab:MethodsAtCommands}.
    \item \textbf{Transport/network-layer metrics:} Downlink throughput via \texttt{iperf3} (\ac{TCP}) and round-trip-time via \ac{ICMP} pings at \qty{1}{\hertz}.
\end{itemize}

Multiple samples were taken at each waypoint to mitigate stochastic variability, and streams were synchronized by nearest-timestamp matching.
Randomly selected throughput samples were sanity-checked against theoretical maxima for the network configuration, based on \cite[Sec.
4]{3GPP2023UserEquipmentRadioAccessCapabilities} to detect gross anomalies.

\begin{table}[h]
    \centering
    \footnotesize
    \caption{AT commands used for collecting radio metrics from the Quectel RM500Q-GL modem. For details, see \cite{Quectel2021ATCommandsManual}.}
    \label{tab:MethodsAtCommands}
    \begin{tabular}{ll}
        \toprule
        \textbf{AT command} & \textbf{Query} \\
        \midrule
        \texttt{AT+QENG} & Primary serving cell information \\
        \texttt{AT+QSINR} & \Acl{SINR} \\
        \texttt{AT+QRSRP} & \Acl{RSRP} \\
        \texttt{AT+QRSRQ} & \Acl{RSRQ} \\
        \texttt{AT+QNWCFG=nr5g\_csi} & \Acl{CSI} \\
        \texttt{AT+QNWCFG=down} & Average downlink throughput \\
        \bottomrule
    \end{tabular}
\end{table}

To address potential localization drift in the large open space, we added small landmarks and performed occasional manual re-positioning when necessary.
The hall was cleared of all other movable objects to match the 3D model geometry, and measurements were conducted during off-hours to minimize external interference.

The measurement campaign yielded approximately \num{9000} data points throughout the Reactor Hall.
Following data cleansing to remove measurement anomalies, approximately \num{7000} valid samples remained for analysis.
These samples were distributed across approximately \num{900} unique spatial locations.
For each of these locations, we computed statistical summaries including mean, standard deviation, and other distributional parameters.
The complete dataset, including raw measurements, processed statistics, and spatial coordinates, is publicly available~\cite{AlNasrallah2025R1EP5GChannelMeasurement}.

\subsection{Data-Driven Throughput Modeling}
\label{sec:MethodsDataDriven}
Using the collected measurement data, we trained \ac{GP} models to estimate the spatial distribution of downlink throughput.
Inputs $x$ were the 2D spatial coordinates, and the response $y$ was the measured downlink throughput.
Because measurement uncertainty varies spatially---e.g., due to differing sample counts or local conditions---we employ \emph{heteroscedastic} noise modeling by providing the per-point standard deviation term individually for each training location.
%
The measurement dataset was split into \qty{70}{\percent} training and \qty{30}{\percent} test using random sampling; multiple random seeds were used to assess variability across splits.
The regression problem was implemented using the \texttt{scikit-learn} library \cite{Pedregosa2011ScikitlearnMachineLearningPython}.

Three stationary covariance kernels were evaluated: \ac{RBF}, \ac{M1}, and \ac{RQ}; see \cref{sec:BackgroundKernelFunctions} for more details.
For each kernel and random seed, hyperparameters were optimized by maximizing the log marginal likelihood with five random restarts to mitigate local optima.
To improve convergence, hyperparameter initial guesses were derived empirically from each training set:
\begin{itemize}
    \item $\sigma_f^2$: measurement variance across all training data,
    \item $\ell$: median $L^2$-distance to 10th nearest neighbor,
    \item $\sigma_n^2$: median per-point measurement variance.
\end{itemize}
Finally, trained models were used to generate spatial throughput predictions over the same \qty{0.5}{\meter} grid as the simulation for direct comparison.

\subsection{Comparative Analysis Framework}
\label{sec:MethodsComparativeFramework}
All measurements and predictions were aligned in a common coordinate frame via nearest-neighbor matching (threshold: \qty{0.5}{\meter}).
The signed error is defined as $e_i = \hat{y}_i - y_i$, where $y_i$ is the measured throughput and $\hat{y}_i$ the model prediction (positive indicates over-prediction), and $|e_i|$ denotes the absolute error.
Where relevant, the relative error is defined as $e_i^{\text{rel}} = e_i / y_i$ and is reported in percent.
We evaluate performance using a statistical scorecard, grouped into the following categories.

\subsubsection*{Category 1: Bias (Directional Tendency):}
We begin by quantifying systematic directional error---whether predictions tend to be optimistic (over-prediction) or pessimistic (under-prediction).
\begin{itemize}
    \item Median error
    \item Mean over-/under-prediction magnitudes
    \item Over-/under-prediction rates (percent)
    \item Maximum over-/under-prediction magnitudes
\end{itemize}

\subsubsection*{Category 2: Accuracy (Central Tendency of Error Size):}
Having established directional tendencies, we next measure the typical magnitude of discrepancy regardless of sign. 
\begin{itemize}
    \item \Acf{MAE}
    \item \Acf{RMSE}
\end{itemize}

\subsubsection*{Category 3: Variability (Deviations and Tail Risk):}
Building on central-tendency accuracy, we then characterize dispersion and tail risk to assess the reliability and spread of fluctuating conditions.
Tail-focused statistics highlight rare but consequential deviations and support risk-aware decisions and worst-case planning.
\begin{itemize}
    \item Sample standard deviation of signed errors.
    \item Sample standard deviation of absolute errors.
    \item \Acf{MAD}.
    \item Percentiles of absolute error: $P_{90}(|e|)$ and $P_{95}(|e|)$.
\end{itemize}


\subsubsection*{Visualization of Results}
We complement the summary metrics by examining statistical distributions of the signed and absolute errors using histograms, which quantify central tendency, dispersion, skewness, and tail behavior, and make outliers explicit.

In parallel, spatial heatmaps reveal geographically localized error patterns across environmental contexts (e.g., corners, corridors, and occluded areas), supporting qualitative assessment of the operational conditions under which different predictors perform well or poorly.
The scientific color maps \texttt{batlow} and \texttt{vik} from \cite{Crameri2023ScientificColourMaps} are used in this study to prevent visual distortion of the data and the exclusion of readers with color-vision deficiencies.

The subsequent Results section applies this comparative framework to evaluate all three prediction methods—physics-based simulation, data-driven GPR models, and measurements—using the metrics and visualizations defined above.
\section{Results}
\label{sec:Results}

This section presents the empirical and modeling results obtained from three complementary approaches: the physics-based \ac{SBR} simulator, on-site physical measurements, and data-driven \ac{GPR} models.
We first provide a qualitative assessment of the site-wide 2D prediction maps to identify systematic patterns and failure modes, and subsequently present a more comprehensive statistical comparison using error distributions and a quantitative scorecard of summary metrics.
The analysis follows a systematic progression through the following stages:
\begin{itemize}
    \item \cref{sec:ResultsParametric}: Empirical comparison between simulation and measurements, including qualitative error analysis.
    \item \cref{sec:ResultsInvestigating}: Investigation into the physics-based simulator, focusing on link-layer phenomena.
    \item \cref{sec:ResultsDataDriven}: Evaluation of data-driven \ac{GPR} throughput prediction, with kernel-specific comparisons.
    \item \cref{sec:ResultsStatistical}: Statistical comparison across all methods, including error distributions and summary metrics.
    \item \cref{sec:ResultsSummary}: Summary of key findings.
\end{itemize}

We begin with the key operational metric---downlink throughput---and then interpret the result via the link-adaptation variables (\ac{MCS} and spatial layers).
This ordering reflects that throughput is the end-to-end quantity of interest, whereas other simulated metrics explain \emph{why} the simulator succeeds or fails at a given location.

\subsection{Parametric (Physics-Based) Simulation Results}
\label{sec:ResultsParametric}
The simulated throughput field, shown in \cref{fig:ResultsThruputSimulated}, reveals a clear structure driven by the environment's large-scale geometry.
The open central hall exhibits peak rates up to approximately \qty{740}{\mega\bit\per\second} under the configured \qty{80}{\mega\hertz} carrier and nominal 4-layer transmission.
In contrast, obstructed regions---such as the far left corridor behind multiple concrete walls and the right-hand corner near a large wooden structure---display pronounced degradation.
This pattern aligns with expectations: \ac{LOS} areas yield strong performance, while heavily shadowed or penetration-limited zones exhibit substantial attenuation.

\begin{figure}[htb]
    \centering
    \input{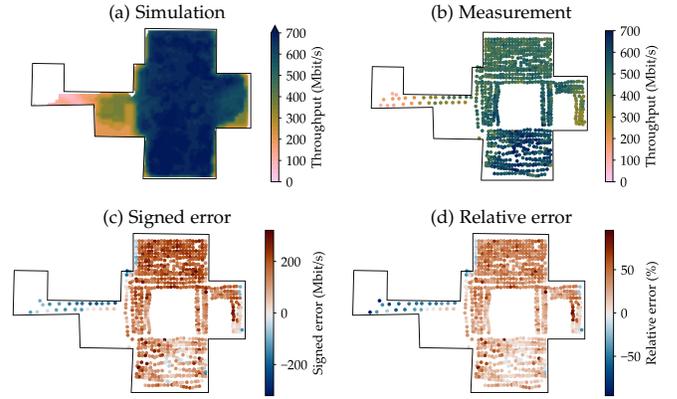}
    \caption{Site-wide 2D map of throughput: simulation vs.\ measurement and associated errors. A positive error indicates over-prediction.}
    \label{fig:ResultsThruputComparison}
\end{figure}

Comparing these simulated predictions against measurements using the nearest-neighbor pairing defined in \cref{sec:Methods} reveals a substantial mismatch.
To quantify the simulator--measurement discrepancy, we compile a scorecard of throughput errors computed over all paired locations; see \cref{sec:Methods}.
Errors are defined so that positive values indicate over-prediction.
\Cref{tab:ResultsErrorMetrics} summarizes bias, accuracy, and variability statistics.

\begin{table}[ht]
    \centering
    \caption{Throughput error scorecard (simulation vs.\ measurement).}
    \label{tab:ResultsErrorMetrics}
    \begin{tabular}
        {l S[table-format=3.1] l}
        \toprule
        \textbf{Metric} & \textbf{Value} & \textbf{Unit} \\
        \midrule
        \textbf{Bias} \\
        Median error & 144.8 & \unit{\mega\bit\per\second} \\
        Mean over-prediction magnitude & 150.9 & \unit{\mega\bit\per\second} \\
        Mean under-prediction magnitude & 74.9 & \unit{\mega\bit\per\second} \\
        Over-prediction rate & 92.4 & \unit{\percent} \\
        Under-prediction rate & 7.6 & \unit{\percent} \\
        Maximum over-prediction magnitude & 321.0 & \unit{\mega\bit\per\second} \\
        Maximum under-prediction magnitude & 221.2 & \unit{\mega\bit\per\second} \\
        \midrule
        \textbf{Accuracy} \\
        \Acf{MAE} & 145.1 & \unit{\mega\bit\per\second} \\
        \Acf{RMSE} & 161.1 & \unit{\mega\bit\per\second} \\
        \midrule
        \textbf{Variability} \\
        \Acf{SD} of signed error & 89.8 & \unit{\mega\bit\per\second} \\
        \Acf{SD} of absolute error & 69.9 & \unit{\mega\bit\per\second} \\
        \Acf{MAD} & 54.6 & \unit{\mega\bit\per\second} \\
        90\textsuperscript{th} percentile of absolute error, $P_{90}(|e|)$ & 232.5 & \unit{\mega\bit\per\second} \\
        95\textsuperscript{th} percentile of absolute error, $P_{95}(|e|)$ & 253.7 & \unit{\mega\bit\per\second} \\
        \bottomrule
    \end{tabular}
\end{table}

As summarized in \cref{tab:ResultsErrorMetrics}, the average error metrics are in the order of approximately \qty{100}{\mega\bit\per\second}, corresponding to roughly \numrange{10}{15}\unit{\percent} relative error, given typical throughput levels in this environment.

\subsection{Investigating Throughput Discrepancies}
\label{sec:ResultsInvestigating}
More importantly, the regional error pattern across the floorplan, visible in \cref{fig:ResultsThruputComparison}, shows a clear trend: agreement is best near the active radio dot (below the central pit), remains reasonable in the far left corridor and right corner, yet degrades above the pit where path length is greatest and multipath complexity increases.

To better understand these discrepancies, we next examine the link-adaptation mechanism.
We first compare the simulator's expected \ac{MCS} with the one actually selected by the radio, i.e., the value reported by the \ac{UE} during the measurement interval.

\subsubsection{Why compare \acs{MCS} rather than \acs{SINR}?}
In the simulator, transitions between \ac{MCS} indices are governed by fixed \ac{SINR} thresholds applied at each sampled location: an \ac{MCS} is selected if the local \ac{SINR} exceeds the minimum threshold for that \ac{MCS}.
In a standards-compliant system, the selected \ac{MCS} reported by the \ac{UE} reflects the system's practical mapping from channel conditions into a stable, usable choice of modulation and coding---capturing feedback, reporting, and short-term adaptation---rather than an instantaneous theoretical optimum based solely on \ac{SINR}.
Because throughput depends directly on the used \ac{MCS} together with the chosen number of spatial layers, a comparison of the \emph{measured} \ac{MCS} provides a more meaningful diagnostic of throughput discrepancies than raw \ac{SINR} values alone.
We therefore begin by evaluating how well the simulator predicts \ac{MCS}, then proceed to examine spatial layer selection.

\subsubsection{\acs{MCS}: Pessimistic but Directionally Correct}
\begin{figure}[htb]
    \centering
    \input{Figures/Results-Comparison-MCS}
    \caption{Site-wide 2D map of \acf{MCS} index (\numrange{0}{27}): simulation vs.\ measurement and associated errors. A positive error indicates over-prediction.}
    \label{fig:ResultsMcsComparison}
\end{figure}

\Cref{fig:ResultsMcsComparison} contrasts the simulated and measured \ac{MCS} fields, revealing a stronger overall agreement.
In the strong-signal region below the pit, \ac{MCS} agrees closely, indicating that the simulator captures the high-performance operating point.
In areas impacted by partial shadowing and complex multipath, the behavior is mixed: the simulator tends to \emph{overestimate} \ac{MCS} in some more distant regions, but it \emph{underestimates} \ac{MCS} substantially in the far left corridor and in the right-hand corner with heavy obstruction.
Overall, despite a mild pessimism bias in difficult regions, the simulated \ac{MCS} tracks the overall structure of the measured \ac{MCS} reasonably well.
Importantly, this suggests that \ac{MCS} prediction (and thus \ac{SINR}) itself is \emph{not} the main driver of the large throughput error reported in \cref{tab:ResultsErrorMetrics}.

\subsubsection{MIMO Spatial Layers: The Main Source of Mismatch}
Having established that \ac{MCS} prediction is largely adequate, we now turn to spatial layer selection---the decisive factor behind throughput mismatch.
The number of \textit{spatial layers} is formally reported by the \ac{UE} as the \acf{RI}, which indicates how many independent data streams (also called data layers) can be transmitted simultaneously over the radio channel.
Higher \ac{RI} values enable greater spatial multiplexing and thus higher throughput, with a maximum of four layers in our test configuration.

It is important to note that in 5G systems, the number of utilized layers \emph{can} be strictly lower than the instantaneous channel rank reported as \ac{RI}.
This occurs when the scheduler limits layers to satisfy low traffic demand, improve power and spectrum efficiency, or mitigate inter-user interference and congestion.
In our study, however, we imposed a sustained downlink load over an isolated \ac{EP5G} data path using a single active radio dot with no other scheduled traffic (see \cref{sec:Methods}).
Given this controlled setup, the likelihood of demand- or congestion-driven layer throttling is exceedingly low, allowing us to interpret observed \ac{RI} values as reflecting genuine channel capacity rather than scheduler constraints.

\begin{figure}[htb]
    \centering
    \input{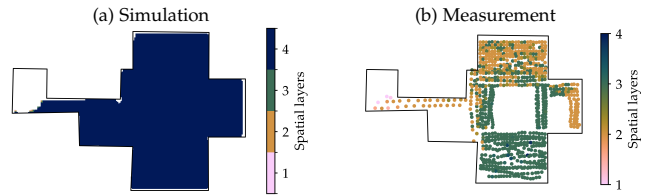}
    \caption{Site-wide 2D map of utilized spatial layers; reported as \acf{RI}.}
    \label{fig:ResultsDataLayerComparison}
\end{figure}

As can be seen in \cref{fig:ResultsDataLayerComparison}, the simulator predicts a \emph{uniform} $\text{RI} = 4$ (four spatial layers) throughout the hall, with only small localized reductions near the far end of the corridor; by contrast, the measured \ac{RI} exhibits clear adaptation:
\begin{itemize}
    \item $\text{RI} \ge 3$ (three spatial layers) in strong-signal \ac{LOS} regions (with only a few locations reaching four layers),
    \item $\text{RI} \approx 2$ (two spatial layers) in further-away regions and under moderate conditions,
    \item $\text{RI} = 1$ (single spatial layer) in heavily attenuated areas (e.g., the far left corridor).
\end{itemize}

Because throughput scales with both spectral efficiency (via \ac{MCS}) \emph{and} spatial multiplexing (via \ac{RI}), overpredicting the available spatial rank inflates the simulated throughput even when the \ac{MCS} is predicted conservatively.
This explains the seemingly paradoxical trend where throughput greatly disagrees in \ac{LOS} regions, despite the predicted \ac{MCS} being close.
Additionally, measurements display non-integer average \ac{RI} at some locations, reflecting rapid rank switching over the sampling window; indicating that the sustainable spatial rank is strongly location-dependent and varies with small-scale fading.

\subsubsection{Summary}
The analysis reveals that the physics-based simulator captures the general structure of indoor propagation---high performance in \ac{LOS} regions, degradation under heavy shadowing---and predicts \ac{MCS} with reasonable fidelity, albeit with conservative bias in the most challenging areas.
However, the dominant error mechanism is \emph{over-prediction of sustainable rank}, which inflates simulated throughput where the real system reduces the number of spatial layers.
This assessment is consistent with the overall throughput error illustrated in \cref{fig:ResultsThruputComparison}: the highest overestimation occurs above the pit, while better agreement emerges near the radio dot and in severely shadowed zones where operational rank is inherently low.

\subsection{Data-Driven Throughput Prediction}
\label{sec:ResultsDataDriven}
Having identified the limitations of the physics-based approach, we now turn to a purely data-driven alternative that predicts downlink throughput directly, for a given 2D location, without explicit modeling of propagation physics.

\begin{figure}[htb]
    \centering
    \input{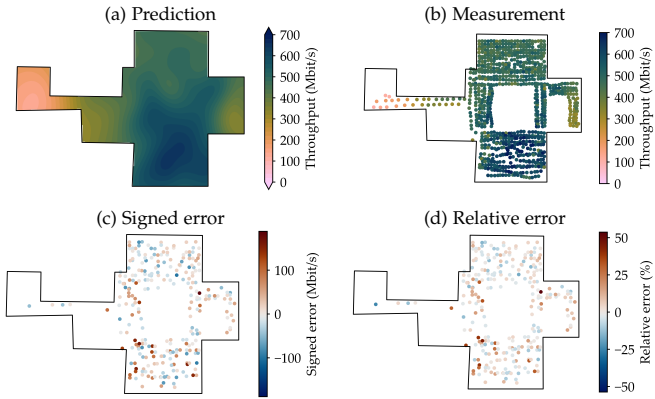}
    \caption{Site-wide 2D map of throughput: \acs{GPR} prediction (\acs{RQ} kernel) vs.\ measurement and associated error.}
    \label{fig:ResultsThruputComparisonRq}
\end{figure}

With the composite kernel formulation described in \cref{sec:BackgroundKernelFunctions}---including the white noise term for regularization---all three kernels produce qualitatively similar predictions that capture both large-scale propagation trends and local variations.
\Cref{fig:ResultsThruputComparisonRq} shows representative \ac{GPR} predictions; the three kernels yield visually comparable maps for mean throughput.

\begin{figure}[htb]
    \centering
    \input{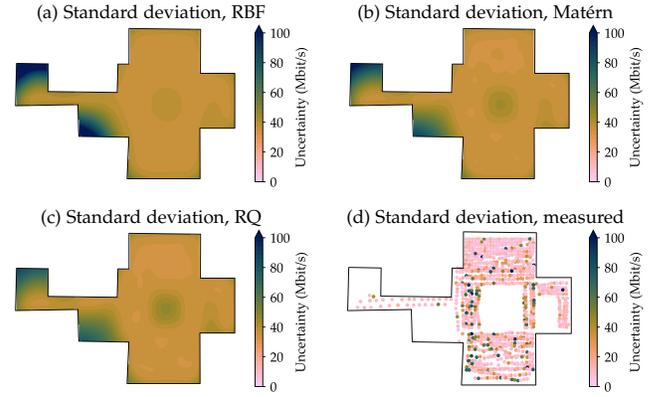}
    \caption{Site-wide 2D map of predictive uncertainty (posterior standard deviation): \acs{GPR} models compared with measured throughput variability. All color maps use the same scale for direct comparability.}
    \label{fig:ResultsGprUncertainty}
\end{figure}

While mean predictions are nearly indistinguishable across kernels, the spatial distribution of predictive uncertainty reveals clear differences in how each kernel represents confidence in unobserved regions.
\Cref{fig:ResultsGprUncertainty} presents a comparative analysis of predictive uncertainty, visualized as posterior standard deviation, across the three kernel functions and contrasted with the empirical throughput variability observed in measurements.

It can be seen that the kernels exhibit modest differences in their ability to differentiate uncertainty spatially:
\begin{itemize}
    \item \textbf{\acs{RBF}:} Relatively uniform posterior variance across the environment, with only slight increase around the central pit and limited localized low-uncertainty regions.
    \item \textbf{\acs{M1}:} Somewhat increased posterior variance in sparsely measured regions (e.g., around the pit), with emerging areas of lower uncertainty that begin to align with reduced measurement variability.
    \item \textbf{\acs{RQ}:} More noticeable spatial differentiation of uncertainty, with moderately elevated variance in unmeasured regions, and slightly larger coherent areas of low predicted uncertainty where empirical measurement variability is consistently low.
\end{itemize}
This progression reflects the kernels' structural flexibility: the \ac{RQ} kernel's scale-mixture formulation better captures the multi-scale spatial structure of indoor propagation, producing uncertainty estimates that more faithfully track the underlying measurement variability.
For applications requiring calibrated uncertainty estimates, such as risk-aware planning, it is clear that the kernel choice becomes relevant; we discuss this further in \cref{sec:Discussion}.

\subsection{Statistical Comparison Across Methods}
\label{sec:ResultsStatistical}
\begin{figure}[htb]
    \centering
    \input{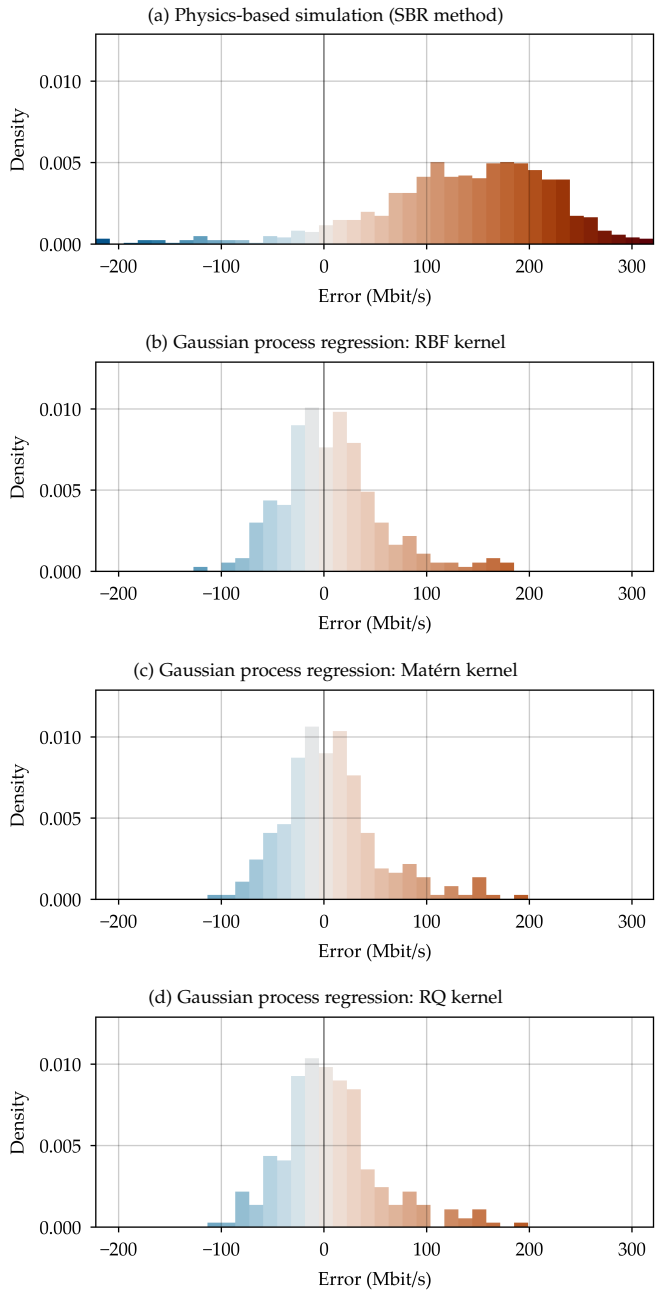}
    \caption{Throughput error histograms across methods, using common bin edges and shared axes. 
    All histograms are PDF-normalized, making shapes directly comparable across datasets of different sizes
    .}
    \label{fig:ResultsThruputHistogram}
\end{figure}

To compare methods in a way that is visually interpretable and statistically sound, we examine the distributions of mean throughput error using per-method histograms with common bin edges and shared axes (\Cref{fig:ResultsThruputHistogram}).
All histograms are normalized as \acp{PDF}, i.e., counts are scaled so that the total area is one.
This normalization removes the visual dependence on sample size and makes the shapes directly comparable across methods:
biases manifest as shifts of mass away from zero, spread reflects variability, and heavy tails indicate outliers.
Because the \ac{GPR} models are trained and evaluated on a \qty{30}{\percent} subsample whereas the simulator is compared against all measured points, it is important to note that PDF estimates from smaller sample sizes may exhibit higher variance and appear noisier, particularly in the tails.
Complementing the visual comparison, \Cref{tab:ResultsErrorMetricsAll} reports a quantitative scorecard that aggregates the same information into interpretable summary statistics.

\begin{table}[ht]
    \centering
    \caption{Throughput error scorecard: all methods.}
    \label{tab:ResultsErrorMetricsAll}
\begin{tabular}
	{l S[table-format=3.1] S[table-format=3.1] S[table-format=3.1] S[table-format=3.1] l}
	\toprule
	\textbf{Metric} & \multicolumn{1}{c}{\textbf{Sim}} & \multicolumn{1}{c}{\textbf{RBF}} & \multicolumn{1}{c}{\textbf{M1}} & \multicolumn{1}{c}{\textbf{RQ}} & \textbf{Unit} \\
	\midrule
	\textbf{Bias} \\
	Median error & 144.8 & 2.5 & 1.5 & 1.7 & \unit{\mega\bit\per\second} \\
	Mean over-pred. & 150.9 & 42.6 & 42.0 & 41.6 & \unit{\mega\bit\per\second} \\
	Mean under-pred. & 74.9 & 30.7 & 29.2 & 29.5 & \unit{\mega\bit\per\second} \\
	Over-pred. rate & 92.4 & 52.2 & 51.5 & 51.9 & \unit{\percent} \\
	Under-pred. rate & 7.6 & 47.8 & 48.5 & 48.1 & \unit{\percent} \\
	Max over-pred. & 321.0 & 180.0 & 186.7 & 188.7 & \unit{\mega\bit\per\second} \\
	Max under-pred. & 221.2 & 114.3 & 102.0 & 100.1 & \unit{\mega\bit\per\second} \\
	\midrule
	\textbf{Accuracy} \\
	\Acs{MAE} & 145.1 & 36.9 & 35.8 & 35.8 & \unit{\mega\bit\per\second} \\
	\Acs{RMSE} & 161.1 & 49.6 & 49.0 & 49.0 & \unit{\mega\bit\per\second} \\
	\midrule
	\textbf{Variability} \\
	\Acs{SD} (signed) & 89.8 & 49.0 & 48.4 & 48.5 & \unit{\mega\bit\per\second} \\
	\Acs{SD} (absolute) & 69.9 & 33.1 & 33.4 & 33.5 & \unit{\mega\bit\per\second} \\
	\Acs{MAD} & 54.6 & 26.8 & 23.9 & 24.4 & \unit{\mega\bit\per\second} \\
	$P_{90}(|e|)$ & 232.5 & 77.6 & 77.9 & 78.6 & \unit{\mega\bit\per\second} \\
	$P_{95}(|e|)$ & 253.7 & 98.0 & 99.0 & 99.5 & \unit{\mega\bit\per\second} \\
	\bottomrule
\end{tabular}
\end{table}

The bias group captures location and asymmetry, the accuracy group reports overall error levels, and the variability group summarizes dispersion and tail behavior.
Together, the figure and the table provide complementary views: distributional shape for intuition and side-by-side numbers for precise comparison.

\subsection{Summary}
\label{sec:ResultsSummary}
Compared with the physics-based simulator, \ac{GPR}-based predictors yield consistently lower error and bias across all metrics: the median error is near zero, \ac{MAE} and \ac{RMSE} are reduced by roughly two-thirds, the 90th and 95th percentiles are markedly smaller, and the worst-case magnitudes are also diminished.
This improvement is visually evident in \cref{fig:ResultsThruputHistogram}, where the \ac{GPR} error distributions are centered near zero with significantly narrower spread, contrasting sharply with the simulator's pronounced positive skew and elongated right tail reflecting systematic over-prediction.

\section{Discussion}
\label{sec:Discussion}
\acresetall

This study set out to evaluate whether physics-based simulation and data-driven modeling can reliably predict \ac{5G} downlink throughput for communication-aware robotic planning in industrial environments.
We now interpret the comparative results to answer RQ1 and RQ2, explain the mechanisms behind observed performance differences, and assess implications for communication-aware planning.

\subsection{Summary of Key Findings}
\label{sec:DiscussionKeyFindings}
We structure our findings around the two research questions posed in \cref{sec:Introduction}.

\subsubsection*{RQ1: Physics-Based Simulation Accuracy}
The physics-based \ac{SBR} simulator successfully captures the spatial structure of indoor \acs{5G} propagation, correctly predicting high \ac{MCS} indices in \ac{LOS} regions with degradation in shadowed areas.
However, it exhibits systematic throughput overestimation: the model consistently over-predicts at \qty{92}{\percent} of measured locations, with a median error of approximately \qty{145}{\mega\bit\per\second}.
The dominant error source is over-prediction of \acs{MIMO} spatial layers; 
the simulator predicts near-uniform four-layer transmission, while measurements reveal substantial adaptation.
Generally, we see three or more layers only in strong \ac{LOS} regions, two layers in moderate conditions, and single-layer fallback in heavily attenuated areas.
This discrepancy directly inflates predicted throughput, as it scales multiplicatively with spectral efficiency and spatial multiplexing gain.
Notably, while prior surveys have identified proper \ac{MIMO} modeling as essential for \ac{5G} channels---emphasizing channel-level properties such as spherical wavefront and array non-stationarity~\cite{Wang2018Survey5GChannelMeasurementsModels}---our results demonstrate that even modest single-user \ac{MIMO} configurations (four antennas) exhibit substantial discrepancies between predicted and realized spatial multiplexing gain.
This is a link-layer phenomenon that channel-centric models do not capture.

Three modeling simplifications likely contribute to this optimistic spatial rank prediction:
\begin{enumerate}
    \item The simulator likely employs idealized antenna patterns and simplified beam management, neglecting the vendor-specific trade-offs between analog and digital precoding stages inherent in hybrid beamforming architectures~\cite{Ahmed2018SurveyHybridBeamformingTechniques5G}.
    \item It likely assumes near-ideal inter-stream orthogonality, employing simplified inter-stream interference models that do not fully capture subspace overlap and interference arising in realistic indoor multipath~\cite{Bojovic2024MIMONetworkSimulatorsDesignImplementation}.
    \item It does not account for the closed-loop feedback and scheduling logic that jointly adjusts \ac{MCS} and \ac{RI} under instantaneous channel and load conditions; accurate modeling of PMI and RI reporting mechanisms can significantly affect predicted performance~\cite{Bojovic2024MIMONetworkSimulatorsDesignImplementation}.
\end{enumerate}

\subsubsection*{RQ2: Data-Driven \acs{GPR} Performance}
The \ac{GPR} models, when properly configured, achieve substantially better predictive performance.
When looking at the error, relative to the simulator, \ac{GPR} achieves:
\begin{itemize}
    \item elimination of bias (median error near zero, close to 50--50 split between over- and under-prediction);
    \item approx. two-thirds reduction in \acs{MAE} and \acs{RMSE};
    \item up to \qty{60}{\percent} reduction in the 90\textsuperscript{th} and 95\textsuperscript{th} percentiles;
    \item approx. \qty{25}{\percent} reduction in maximum error magnitudes.
\end{itemize}
One key factor enabling this performance is proper noise modeling through the white noise term in the composite kernel formulation.
When omitted, all kernels exhibit underfitting of large-scale trends while overfitting local variations.
With proper noise modeling, the three evaluated kernels produce statistically indistinguishable mean predictions---kernel choice becomes secondary to noise regularization for point prediction accuracy.

However, kernels differ in their posterior uncertainty structure, though the effect is subtle.
Broadly speaking, the \ac{RBF} kernel yields nearly uniform posterior variance, whereas the \ac{RQ} kernel produces variance fields that spatially track empirical measurement variability to a larger degree.
For risk-aware planning applications that incorporate uncertainty bounds, this distinction may become practically important.

\subsection{Implications for Communication-Aware Planning}
\label{sec:DiscussionPlanningImplications}
Both methodologies produce spatially structured predictions suitable for integration with planning algorithms, but exhibit fundamental limitations that persist even under near-ideal experimental conditions.

Physics-based simulation provides physics-grounded extrapolation across the entire environment, including unmeasured regions.
However, the substantial prediction errors---despite near-laboratory conditions with detailed geometry, controlled environment, and no external interference---reveal inherent limitations for end-to-end throughput prediction.
The systematic spatial rank over-prediction stems from the modeling simplifications identified in \cref{sec:DiscussionKeyFindings}, and these limitations are unlikely to be resolved in the near future.
First, accurately modeling hybrid beamforming requires capturing the intricate trade-offs between analog and digital precoding stages, RF chain constraints, and beam management procedures that vary across vendors and deployments~\cite{Ahmed2018SurveyHybridBeamformingTechniques5G}---information that is rarely disclosed.
Second, realistic inter-stream interference modeling demands computationally intensive matrix operations and receiver algorithms (e.g., MMSE-IRC) that many ray-tracing simulators simplify or omit entirely~\cite{Bojovic2024MIMONetworkSimulatorsDesignImplementation}.
Third, closed-loop CSI feedback mechanisms---including PMI and RI selection algorithms---are proprietary, implementation-specific, and interact with real-time scheduling decisions that cannot be replicated without access to vendor firmware~\cite{Bojovic2024MIMONetworkSimulatorsDesignImplementation}.
Together, these factors create a persistent gap between simulated and realized spatial multiplexing gain that per-deployment calibration cannot fully bridge without undermining the primary advantage of simulation-based prediction.
In realistic industrial deployments with incomplete geometry, dynamic obstacles, and time-varying interference, physics-based predictions may exhibit substantially larger deviations.

These findings also temper assumptions underlying much prior work on communication-aware motion planning.
Existing approaches---including resilient planners~\cite{Caccamo2017RCAMPResilientCommunicationawareMotionPlanner}, motion--communication co-optimization~\cite{Ali2019MotioncommunicationCooptimizationCooperativeLoadTransfer}, and radio-map-based indoors planning~\cite{Mu2021IntelligentReflectingSurfaceEnhancedIndoor}---typically treat \ac{SINR} or received signal strength as the primary optimization objective, implicitly assuming that favorable channel conditions translate reliably into high throughput.
Our results challenge this channel-centric paradigm: the physics-based simulator achieved reasonable \ac{SINR} (and thus \ac{MCS}) predictions, yet throughput errors remained substantial due to spatial rank over-prediction.
This suggests that planners optimizing purely for \ac{SINR} may fail to deliver expected throughput in practice, particularly in \ac{MIMO} systems where spatial multiplexing gain depends on hybrid beamforming architectures, inter-stream interference suppression, and closed-loop rank adaptation---none of which are captured by channel-level metrics.
Because the underlying modeling challenges are structural rather than merely a matter of calibration or improved channel estimation, this gap is unlikely to be closed through incremental refinements to physics-based simulators.
For robust communication-aware planning, end-to-end performance prediction---rather than channel-level proxies---should be the target metric.

While data-driven methods offer an alternative path toward end-to-end throughput prediction, they introduce their own operational challenges which have to be addressed.
\Ac{GPR} achieves lower error by learning system behavior directly from measurements, but has cubic-time computational complexity that becomes costly as measurement campaigns scale.
More critically, \ac{GPR} models are conservative and slow to adapt: regions with strong historical measurement density exhibit high confidence in outdated predictions even after physical modifications, requiring extensive remeasurement to override prior beliefs.
This may create an unsustainable operational burden as the environment or network configuration evolves.
Neither approach therefore provides a fully satisfactory solution for reliable throughput prediction under realistic operational conditions.

\section{Conclusion}
\label{sec:Conclusion}
This study compared physics-based and data-driven approaches for predicting \ac{5G} downlink throughput in the KTH Reactor Hall---an underground facility equipped with an \acf{EP5G} network and edge-computing testbed.
A robotic platform collected spatially distributed measurements to validate predictions from a ray-tracing simulator as well as \acf{GPR} models with three distinct kernel types.
The measurement dataset is publicly available to support reproducibility and future research~\cite{AlNasrallah2025R1EP5GChannelMeasurement}.

The central finding challenges a common assumption in communication-aware robotics: that favorable channel conditions---as captured by \ac{SINR} or received signal strength---translate reliably into high throughput.
The physics-based simulator achieved reasonable \ac{MCS} predictions yet systematically over-predicted throughput due to optimistic \ac{MIMO} spatial rank assumptions.
This indicates that planners optimizing channel-level metrics may not achieve expected throughput in practice, highlighting the need for end-to-end performance prediction rather than intermediate channel proxies.
The \ac{GPR} models reduced prediction error by approximately two-thirds and eliminated systematic bias, demonstrating that data-driven approaches can capture end-to-end system behavior that physics-based simulators miss.
Notably, all three kernels produced statistically indistinguishable mean prediction accuracy; however, they differed substantially in their posterior uncertainty structure, with the \ac{RQ} kernel best tracking empirical measurement variability.
Both approaches have practical trade-offs: 
physics-based simulation requires extensive calibration and access to proprietary link-adaptation logic, while data-driven methods require substantial measurement data and adapt slowly to environmental changes.

Future work should explore hybrid approaches combining physics-based priors with data-driven refinement, investigate scalable machine learning methods for larger deployments, and develop online adaptive methods that update predictions during robot operation.
For risk-aware planning applications, attention should be given to covariance kernel selection, as it determines the degree with which the posterior variance aligns with empirical data variability.
Integration with intent-driven autonomous network management and digital network twin frameworks may offer promising directions for bridging the gap between prediction and reliable planning.

\printbibliography

\end{document}